\documentstyle[aps,pre,float,epsf,amsfonts,amssymb]{revtex}
\newcommand{\bu}{{\bf u}}
\newcommand{\bv}{{\bf v}}
\newcommand{\bk}{{\bf k}}

\newcommand{\br}{{\bf r}}
\newcommand{\bx}{{\bf x}}
\newcommand{\fu}{{\bf f}^{\rm u}}
\newcommand{\fv}{{\bf f}^{\rm v}}
\newcommand{\Pt}{{\bf P}^{\rm T}}
\newcommand{\Pl}{{\bf P}^{\rm L}}
\newcommand{\rF}{{\rho}_{\rm F}}

\begin{document}
\draft
\twocolumn[\hsize\textwidth\columnwidth\hsize\csname@twocolumnfalse%
\endcsname
\title{The response function of a sphere in a viscoelastic
two--fluid medium}
\author{Alex J.~Levine and T. C.~Lubensky}
\address{Department of Physics and Astronomy, University of Pennsylvania,
Philadelphia, PA 19104}

\date{\today}
\maketitle
\begin{abstract}
In order to address basic questions of importance to microrheology, we 
study the dynamics of a rigid sphere embedded in a model viscoelastic 
medium consisting of an elastic network permeated by a viscous fluid.
We calculate the complete response of a single 
bead in this medium to an external force and compare the result to the 
commonly--accepted, generalized Stokes--Einstein relation (GSER).  
We find that our response function is well approximated by the GSER 
only within a particular frequency range determined by the material parameters 
of both the bead and the network. We then discuss the relevance of this 
result to recent experiments.  Finally we discuss the approximations made in 
our solution of the response function by comparing our results to the exact 
solution for the response function of a bead in a viscous (Newtonian) fluid.
\end{abstract}
]
\section{Introduction}
Microrheology\cite{Mason:95}
 has become an important experimental probe of the mechanical properties of
soft materials such as actin or other 
bio-polymers\cite{Mackintosh:99,Schnurr:97}.  
It is a class of experimental techniques that measure the response of probe
particles to external forces.  It is generally accepted that the response 
functions measured in microrheology experiments determine the same material
properties as do traditional rheology experiments.
In other words, one expects that it is possible to express the 
response function 
measured in microrheology in terms of the complex shear modulus, $G(\omega)$. 
This connection between the measured response function and $G(\omega)$ 
allows one to obtain rheological data for materials that 
cannot be produced in large quantity or to study the local 
rheological properties of rheologically inhomogeneous materials.  An important 
example of a system which satisfies both of the above criteria 
is the living cell.  Microrheology, therefore, promises to open an 
new window on cellular biology\cite{Yamada:00,Bausch:99}.

There are currently two classes of techniques used to measure probe--particle
responses.  In the active technique, probe particles are subjected to an 
external force ({\it e.g.\/} magnetic\cite{Bausch:99,Magnetic,Amblard:96} or 
laser tweezers\cite{Helfer:00}), and their 
displacements are measured with the aid of microscopes and imaging technology.
In the passive technique, thermally fluctuating positions of particles are 
measured either via direct observation\cite{Crocker:00} or via light 
scattering\cite{Mason:95,Schnurr:97,Mason:97} and  the response function is 
then determined with the aid of the fluctuation--dissipation theorem.
In either case we emphasize that
there is an essential role to be played by theory to establish the connection
between the measured response function and the underlying material 
properties of the medium being investigated.
It is generally assumed\cite{Mason:95,Schnurr:97} that this connection is 
provided by the generalized Stokes--Einstein relation (GSER) in
which the position $\br (\omega)$ of the probe particle (of radius $a$) 
as a function of frequency is given by
\begin{equation}
\label{GSER}
\br (\omega) = \frac{ 1}{6 \pi a G(\omega )} \, {\bf F}(\omega ),
\end{equation}
where ${\bf F}(\omega )$ is the applied force on the particle and $G(\omega)$ is
the complex shear modulus. This result
is the natural generalization of the Stokes mobility of the a rigid, spherical 
particle in a viscous fluid where the complex shear modulus reduces to
$G(\omega ) = -i \omega \eta $.  One can certainly 
measure the mobility of such a spherical probe particle of known radius 
in a Newtonian fluid by observing its Brownian fluctuations (passive mode)
or by a sedimentation experiment 
(active mode) and thereby determine the viscosity of the 
medium\cite{active-passive}.  The generalization of this result embodied by the
GSER [Eq.~(\ref{GSER})] suggests that the analogous experiments performed 
in an arbitrary viscoelastic material will allow one to similarly 
obtain $G(\omega )$ for that material by the application of Eq.~(\ref{GSER}).  
 
In this paper we examine the validity of the GSER through detailed 
calculations of the response of a rigid, spherical probe particle in a 
model viscoelastic medium. In particular, we study a 
two--fluid model\cite{Brochard:77} of a generic viscoelastic medium 
in which a viscoelastic network is viscously coupled to a permeating fluid. This
model, which we study in a continuum limit, may be taken to represent a 
gel or an uncrosslinked polymer solution studied at frequencies larger than 
its plateau frequency.   

There are two basic reasons to question the validity of the GSER: First, the 
mode structure of a multi-component medium is more complex than that of a simple
fluid.  A probe particle moving at frequency $\omega$ will excite modes other
than simple shear modes, and its response to external forces will, in general,
depend on all of these modes in a way not simply described by $G(\omega)$.
Second, at frequencies accessible to microrheology experiments, which are much
greater than those accessible to traditional rheology experiments, effects of 
the inertia of both the particle and the medium\cite{longtime}, which are not 
included either in the simple Stokes--Einstein relation or in the GSER, may
be important.  We will investigate both of these effects.

The fundamental results of this work have already been presented 
elsewhere\cite{Levine:00}; here we elucidate the details of our 
approximate calculational scheme as well as provide a further discussion 
of the results.  The remainder of this paper is organized as follows: In 
section \ref{medium} we discuss the basic two--fluid model in some detail, 
describing the hydrodynamic modes of the system.  In section \ref{ball-medium}
we describe the approximate calculation of the response of a rigid, spherical
particle embedded in such a medium.  Using a comparison of the results of our 
technique to the well-known result for the drag on
an oscillating sphere in a viscous fluid\cite{Landau}, we discuss the range of
validity of our approximation.  In section \ref{response} we determine 
the requisite conditions for the response function to be 
approximated by the GSER and detail the cross--over to non--GSER like 
behavior as a function of frequency. 
Finally we conclude in section \ref{conclusions} with a discussion of the 
limits of the validity of the GSER in which we apply our results to the 
experiments of Schnurr {\it et al.\/}\cite{Schnurr:97}.  In addition we 
compare the importance of
inertial effects in traditional rheology to those in microrheology.  
The full calculation of the inertial effects upon a traditional, 
parallel--plate, rheological measurement applied to our two-fluid model is 
presented in the appendix. 

\section{The Two--Fluid Medium}
\label{medium}

\begin{figure}
\epsfxsize=0.95\columnwidth
\centerline{\epsfbox{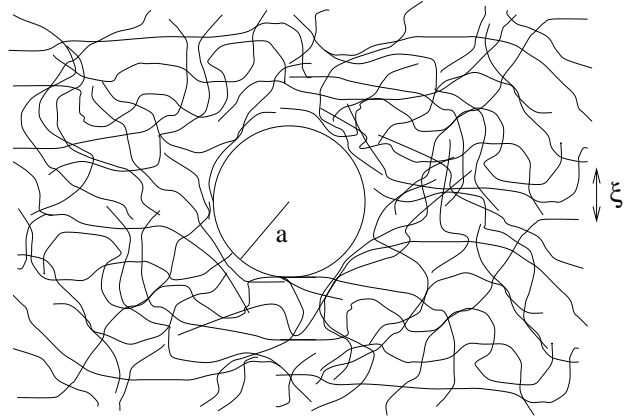}}
\vspace{0.1cm}
\caption[]{
Microscopic model of the two--fluid medium.  The viscoelastic network is
represented by the lines in the figure.  The network has a characteristic
mesh size given by $\xi $.  The probe particle is a sphere of radius $a$ shown
in the center of the figure.  The background viscous fluid (not shown)
permeates the network.
}
\label{picture_one}
\end{figure}

Our model viscoelastic medium consists of a viscoelastic network 
characterized by a displacement variable $\bu$ that is viscously coupled 
via a friction coefficient $\Gamma $  to an incompressible, Newtonian 
fluid characterized by a velocity
field $\bv$.  (See Fig.\ \ref{picture_one}.)  
The viscoelastic network with mesh size $\xi $ is macroscopically isotropic
and homogeneous.  At length scales larger than $\xi $, it is characterized 
by an isotropic continuum elasticity with shear and bulk Lam\`{e} 
coefficients $\mu$ and $\lambda$, which may in general be complex functions
of frequency $\omega$.
Because of the 
viscous coupling of the elastic network to the fluid, there is a drag force
density acting on the network due to its motion through the fluid, in addition
to the force densities resulting from the local, network strain field.

The linearized equation of motion for the displacement field in the presence 
of an externally imposed force density $\fu$ acting directly on the network is
\begin{equation}
\label{ueom}
\rho \ddot{\bu} - \mu \nabla^2 \bu - \left( \lambda + \mu \right) \nabla 
\left( {\bf \nabla } \cdot \bu \right) = - \Gamma \left( \dot{\bu} - 
\bv \right) + \fu, 
\end{equation}
where $\rho$ is the mass density of the network.  The frictional force density
in the above equation is proportional to the local {\it relative\/} velocity
of the network and the background fluid as is required by Galilean invariance.
We may estimate the viscous coupling constant $\Gamma $ as follows:  If a 
strand of the network of length equal to the characteristic mesh size $\xi$
moves relative to background fluid at a velocity $\bv$, the drag force 
it experiences is approximately $\eta \xi  \bv$ where $\eta$ is the fluid
viscosity.  The drag force density on the network, given by $\Gamma \bv$ in 
Eq.~(\ref{ueom}) above, is then $\eta \xi \bv/ \xi^3 \sim 
\eta \bv / \xi^2$.  We then determine that $\Gamma \sim  \eta/\xi^2$.   

The fluid velocity field $\bv$ obeys the linearized, 
incompressible Navier-Stokes equation
with viscosity $\eta$.  Including the drag of the network upon the fluid, we
find
\begin{eqnarray} 
\label{veom}
\rho_{\rm F} \dot{\bv} - \eta \nabla^2 \bv + {\bf \nabla } P &=& \Gamma \left( 
\dot{\bu} - \bv \right) + \fv \\
\label{incompress}
{\bf \nabla } \cdot \left[ \left( 1 - \phi \right) \bv + \phi \dot{\bu} \right] 
&=& 0,
\end{eqnarray} 
where $\rho_{\rm F}$ is the mass density of the fluid, $\phi$ is the volume 
fraction of the elastic network,  and  $P$ is the pressure.
 $\fv$, in analogy to $\fu$, is an externally imposed force density acting 
on the fluid. We note that Eq.~(\ref{incompress}) 
demands the incompressibility of the 
total solution rather than that of the solvent (background fluid) alone. 
However, as we are primarily interested in discussing 
microrheological experiments on stiff bio-polymers such as actin, which form
entangled solutions at extremely low volume fractions\cite{lowphi}, we may
assume that $\phi \ll 1$. In this limit Eq.~(\ref{incompress}) becomes 
the standard 
condition of the incompressibility of the background fluid, ${\bf \nabla} \cdot 
\bv \simeq 0$.  Similarly, because of the sparseness of the net we may assume 
that $\rho (= \phi \rho_{\rm net})   \ll \rho_{\rm F} (=
(1 - \phi) \rho_{\rm fluid}) $ where $\rho_{\rm net}$ and $\rho_{\rm fluid}$
are, respectively, the densities of the pure network and solvent.  Because 
of this inequality we will later be able to simplify our results by making the 
reasonable approximation that $\rho \simeq 0$.  

Lastly we comment on the 
validity of the linearization of the Navier-Stokes equation in Eq.~(\ref{veom}).
In order for our linearization to be valid, the force density associated with 
the convective term $ \rho \bv \cdot {\bf \nabla } \bv$ omitted from the 
linearized Navier--Stokes equation coupled to $\bu$ [Eq.~(\ref{veom})]  must be 
small compared to the other force densities in the system.  To get a sense 
of when this term is unimportant, we investigated excitations at frequency 
$\omega$ in the large $\omega \Gamma$ limit in which the effective linearized
velocity equation becomes 
\begin{equation}
\rF \dot{\bv} - \frac{G(\omega)}{-i \omega} \nabla^2 \bv = 0,
\end{equation}
where $G(\omega) = \mu - i \omega \eta$.  
Thus the convective term can be neglected provided $ \bv \cdot {\bf \nabla}
\bv \ll 1/(-i \omega) [ - \omega^2 + G(\omega)/\rF ] \bv $.
Considering the 
harmonic motion of the particle with frequency $\omega$ and amplitude $\ell$,
we find that there are two cases to study.
First, at low frequencies the second term on the RHS of the
above equation  dominates over the first, and the linearization 
of the Navier-Stokes equation requires that 
$\omega^2 a \ell \rho /G(\omega) \ll 1$, which is just the condition that the 
Reynolds number be small.  At higher frequencies, however, the first or 
inertial term on the RHS of the above equation dominates over the second, and 
the linearization now must be based on the inequality, 
$\ell/a \ll 1$\cite{Landau}.  
Thus, the analysis presented in this paper can be extended to 
high frequencies where the inertial terms dominate the response function and 
the particle Reynolds number is high, provided 
the amplitude of the probe particle's oscillation is small compared to 
its own radius. 
Similar conclusions follow when $\omega \Gamma $ is not large.  
The ability to model high--frequency
dynamics is important in the study of the response function since microrheology
allows the experimenter, in principle, to probe response functions at very high
frequencies.  Current experiments, however, have not yet explored the mechanical
response of beads at high enough frequencies to leave the 
low--Reynolds--number regime.  

We now discuss the hydrodynamic modes of the two--fluid medium.  An equilibrium
crystal in a one--component fluid background ({\it e.g.\/} a colloidal solid)
has nine hydrodynamic modes\cite{Pershan:72,Lubensky:book} (modes with 
frequencies vanishing with wavenumber) arising from three broken translational
symmetries and the conservation of two masses, energy, and momentum.  There are,
one heat diffusion, one relative mass diffusion, one vacancy diffusion, two 
longitudinal sound, and four transverse sound modes.  In our model, network
crosslinks are rigidly fixed so there is no vacancy diffusion.  Our system 
is incompressible so there are no longitudinal sound modes.  In addition, 
we ignore heat diffusion.  We, therefore, expect our model two--fluid system
to have five hydrodynamic modes: one relative mass diffusion mode and four 
transverse sound modes.  Our fundamental dynamical equations 
[Eqs.~(\ref{ueom})-- (\ref{incompress})] contain three $\bu$ equations with two
time derivatives and two independent $\bv$ equations with one time derivative,
have a total of eight modes.  We therefore expect to find three 
non-hydrodynamic modes in addition to the five hydrodynamic modes.

We now determine the modes of the two-fluid model by 
Fourier transforming Eqs.~(\ref{ueom})--(\ref{incompress}) with
$\phi$ and the external forces $\fu, \fv$ set to zero.  After eliminating 
the pressure $P$ using the incompressibility of the fluid we find
\begin{eqnarray}
\label{fourier1}
\left[\Delta^{\rm T} (\bk, \omega ) \Pt_{\alpha \beta } +  
\Delta^{\rm L}(\bk, \omega )
\Pl_{\alpha \beta } \right] u_\beta - \Gamma v_\alpha &=& 0\\
\label{fourier2}
i \omega \Gamma \Pt_{\alpha \beta } u_\beta + \Pi(\bk, \omega ) \Pt_{\alpha
\beta } v_\beta &=& 0\\
\label{fourier3}
\Pl_{\alpha \beta } v_{\beta } &=& 0
\end{eqnarray}
where we have defined $\Delta^{\rm T}(\bk, \omega ) = 
- \omega^2 \rho + \mu k^2 - 
i \omega \Gamma $, $\Delta^{\rm L}(\bk, \omega ) = - \omega^2 \rho + 
( 2 \mu + \lambda ) k^2 - i \omega \Gamma$, and $\Pi(\bk, \omega ) = -i \omega
\rho_{\rm F} + \eta k^2 + \Gamma$.  We have also introduced the standard 
transverse ($\Pt(\bk) = \delta_{\alpha \beta } - \hat{k}_\alpha 
\hat{k}_\beta$) and longitudinal 
($\Pl_{\alpha \beta }(\bk) = \hat{k}_\alpha \hat{k}_\beta$)
 projection operators.  The condition for nontrivial solutions for 
$\bu$ and $\bv$ gives the following result:
\begin{equation}
\label{determinant}
\Delta^{\rm L}(\bk, \omega ) \left[  i \omega \Gamma^2 + 
\Delta^{\rm L}(\bk, \omega ) \Pi(\bk, \omega ) \right]^2 = 0.
\end{equation}
The first factor on the left hand side of Eq.~(\ref{determinant})
is quadratic in $\omega$ while the second factor, which is cubic in $\omega $,
is squared so that the total expression is an eighth order 
polynomial in $\omega $.
Its roots, which correspond to the modes of the system, are 
clearly divided into two sets.  The first set, which are roots of the first 
factor on the LHS of Eq.~(\ref{determinant}), consists of two 
longitudinal modes. The second set, coming from the 
roots of the second factor in 
Eq.~(\ref{determinant}), represent the remaining six transverse modes.  These 
transverse modes come in three identical pairs corresponding to the two 
possible polarization states of the transverse waves.  We first consider the
transverse modes in more detail.

Ignoring the polarization--state degeneracy for the moment, 
two of the three transverse modes are a pair of propagating shear waves in 
the medium.  For small $k$, the dispersion relation is given by
\begin{equation}
\label{dispersion-shear}
\omega ( \bk ) = \pm \sqrt{\frac{ \mu}{\rho + \rho_{\rm F} }} k  - \frac{ i}{2}
\left[ \frac{ \mu \rF^2}{\Gamma \left( \rho + \rF \right)^2 } + 
\frac{ \eta }{\rho + \rho_{\rm F} } \right]  k^2.
\end{equation}
Counting polarization states, these shear modes constitute four of the 
five hydrodynamic modes of the system.
The phase velocity is similar to that of the transverse sound modes in the 
elastic medium when decoupled from the background fluid.  The shear wave speed
of the two--fluid model is identical to that of a one--component elastic 
solid with mass density replaced by the total combined network -- fluid mass
density $\rho + \rho_{\rm F}$.
The effect of the fluid coupling can be seen in the damping rate of this mode. 
In the weak coupling limit, where $\Gamma $ is small, the dominant 
contribution to the damping rate comes from the first term in the brackets 
which arises from the relative motion of the elastic network against the 
background fluid.  The strong coupling limit, on the other 
hand, derives its damping from viscous dissipation 
(the second term in brackets) in the fluid 
which, in this limit, is sheared as it moves with the network.  It 
should be noted that the $\Gamma \longrightarrow 0$, decoupled  limit is not 
easily apparent in the above result.  The long wave length approximation has 
been used in the above derivation which corresponds to taking 
$k \ll \sqrt{\Gamma/\eta} \sim 1/\xi $, where the last expression of the right 
hand side was produced using our estimate, $\Gamma \sim \eta/\xi ^2$. 

The third transverse mode  has a 
finite decay rate at zero wavevector and corresponds to a relative motion of 
the network and fluid that comprises our two--component medium.  To lowest order
in wavevector its decay rate is given by
\begin{equation}
\label{trans-slosh}
\omega ( \bk ) = -i \Gamma \left( 
\frac{ \rho_{\rm F} + \rho}{\rho_{\rm F} \rho} \right) 
\end{equation}
where we have dropped ${\cal O}(k^2)$ corrections to the damping rate.  

To examine the longitudinal modes of the medium it is convenient to allow the
fluid to have a finite compressibility $\chi^{-1} = \rF \partial P / \partial 
\rho_{\rm F}$ at first and then to take the incompressible limit later.
We, therefore, introduce a variable fluid density 
via $ \rho_{\rm F} \longrightarrow \rho_{\rm F} + \delta \rho$, that obeys the
equation of motion: $\dot{\delta \rho} = - \rF {\bf \nabla } \cdot \bv$.  
Projecting out only the longitudinal degrees of freedom of the system, we have
to lowest order in wavevector a pair of propagating longitudinal sound 
modes with dispersion relations of the form:
\begin{equation}
\label{sound-long}
\omega_{\rm sound} (\bk ) = \pm \sqrt{ \frac{ (2 \mu + \lambda ) 
+ \chi \rF}{\rho + \rF}} k  - i d(\chi) k^2,
\end{equation}
where $d(\chi) k^2$ is the decay rate of the sound modes,
and two over--damped modes with decay rates given by
\begin{eqnarray}
\label{long-damped}
\omega_{\rm long} ( \bk ) &=& - i \frac{ (2 \mu + \lambda ) \rF \chi}{\Gamma \left( 
2 \mu + \lambda + \rF \chi \right) } k^2, \\
\label{long-gapped}
\omega ( \bk ) &=& - i \frac{ \Gamma \left( \rho + \rF \right) }{\rho \rF}.
\end{eqnarray}
We note the mode with a finite decay rate at $k = 0$ [Eq.~(\ref{long-gapped})] 
has an identical dispersion
relation to the gapped transverse mode, Eq.~(\ref{trans-slosh}).  We, 
therefore, identify it as the 
longitudinal counterpart to the transverse modes in which there is relative 
motion between the network and the fluid.  In the incompressible limit 
($\chi \longrightarrow \infty$), the longitudinal sound velocity becomes 
infinite as expected.  The decay rate diverges as well, 
$ d(\chi) \longrightarrow \infty$, thus these two modes (Eq.~\ref{sound-long})
in which the network {\it and\/} fluid experience compressions and rarifactions
do not concern us in the incompressible limit.  What will be more 
interesting with regard to the calculation of the response of the bead is 
the fifth hydrodynamic mode (Eq.~(\ref{long-damped}))
whose decay rate remains finite in the 
incompressible limit.  In that limit we find that the decay rate takes the 
form
\begin{equation}
\label{long-damped-incom}
\lim_{\chi \longrightarrow \infty} \omega_{\rm long} (\bk ) = -i \frac{ 2 \mu +
\lambda }{\Gamma } k^2.
\end{equation}
This mode is the relative mass diffusion mode of the system.  The network 
density (described by $\delta \rho/\rho = - {\bf \nabla} \cdot \bu$) changes
and relaxes diffusively while that of the background fluid remains fixed.  
The existence of a slowly decaying longitudinal mode not present in an 
incompressible viscous fluid has consequences for the validity of the GSER 
in our two--fluid model. 

\section{Calculating the Response Function}
\label{ball-medium}

To calculate the response of the probe particle to an applied force, we will
need to introduce the rigid probe particle into the two--fluid medium described
in the previous section.  The complete solution of the problem requires that 
one solve Eqs.~(\ref{ueom})--(\ref{incompress}) with time 
derivatives replaced by $-i \omega $, $\fu$, $\fv$ set equal to zero, and 
the enforcement of the 
correct boundary conditions at the surface of the probe sphere, {\it i.e.\/},
\begin{equation}
\label{boundary_con}
\bu \left( \left| {\bf x} \right| = a, \omega \right)  = 
\bv \left( \left| {\bf x} \right| = a, \omega \right) /-i \omega = 
{\bf r}(\omega ), 
\end{equation}
where ${\bf r}(\omega )$ is the frequency--dependent 
position of the center of the probe sphere. In addition we would need to apply
the boundary condition that both $\bu$ and $\bv$ go to zero far from the 
sphere.  After calculating the displacement and velocity fields, 
 $\bu( {\bf x}, \omega )$ and $\bv( {\bf x}, \omega )$, that solve this
boundary value problem,
we would then calculate, the force ${\bf F}_{\rm b}$ exerted on the sphere 
by the medium by integration of the appropriate components 
of the stress tensor over the surface of the probe sphere. 
Newton's second law applied 
to the probe sphere (of mass $M$) under the influence of the externally applied
force ${\bf F}(\omega )$,
\begin{equation}
\label{Newt}
-\omega^2 M {\bf r}(\omega ) - {\bf F}_{\rm b}(\omega ) = {\bf F} ( \omega ),
\end{equation}
leads to a determination of the response function $\alpha(\omega)$, where
$\alpha$ is defined by
\begin{equation}
\label{response_function}
{\bf r}(\omega ) = \alpha (\omega ) {\bf F}(\omega ).
\end{equation}
The calculation outlined above is possible for the case of a 
simple viscous medium
but becomes more difficult for the two--fluid medium we study.  We 
will, therefore, apply a less rigorous procedure. To justify this approximation
we verify in Appendix \ref{Newtonian_fluid} that our method does, in fact, 
reproduce the correct 
frequency--dependent response function over some finite frequency range in 
the simpler problem of a sphere in a Newtonian fluid.  
Fortunately, the interesting features of microrheological measures can still 
be explored in the frequency range still available to our investigation.

Here we briefly outline our approximate calculation:   
As a first step in this procedure, we restore the
applied force densities $\fu$ and $\fv$ in the equations of motion.  
We then calculate the the displacement field $\bu$ and the fluid 
velocity field  $\bv$ everywhere in the medium as a function of the, as yet 
undetermined values of the two applied force densities, $\fu$ and $\fv$.  These
force densities will be used to represent the forces applied to  the medium
by the probe sphere.  We, therefore, localize these forces at the sphere by
setting
\begin{equation}
\label{force-label}
{\bf f}^{\rm u,v}(\bk, \omega) = {\bf F}^{\rm u,v}(\omega ) 
\Theta \left( k_{\rm max} - \left| {\bf k} \right| \right), 
\end{equation}
where $\Theta(x)$ is the unit step function and $k_{\rm max} = \pi / 2 a$ is
the large wavevector cutoff.  One role of the bead is to cut off the spectrum
of allowed fluctuations of the medium at the length scale of the probe particle
radius.  It it then clear that the wavevector cutoff is proportional to the 
inverse particle radius; the numerical coefficient is chosen to produce the 
correct low--frequency Stokes mobility of the spherical particle in a 
Newtonian fluid as verified in Appendix \ref{Newtonian_fluid}.
We note that this abrupt cutoff in Fourier space cannot be 
strictly valid as we really want a sharp cutoff in the real space, 
applied--force profile. However, we will show that this simple scheme is 
sufficient to reproduce standard hydrodynamic results concerning the 
frequency--dependent response of an oscillating sphere in a viscous fluid 
which we believe justifies our confidence in our more general application 
of the approach. 

Following the procedure outlined above, we now have a solution for the 
motion of the bead in terms of two, as yet 
unknown, forces ${\bf F}^{\rm u}, {\bf F}^{\rm v}$ due to the bead acting on 
the elastic network and viscous fluid respectively.  We determine a relation 
between these two forces by requiring that the boundary condition 
Eq.~(\ref{boundary_con}) at the sphere sphere be satisfied.  The total force
that the bead exerts on the two-fluid medium is therefore given by: 
$- {\bf F}_{\rm b} = - {\bf F}^{\rm u} -  {\bf F}^{\rm v}$.  
We have now calculated
the displacement of the sphere in terms of ${\bf F}_{\rm b}$.  By inverting this
relation and using it in Eq.~(\ref{Newt}) can calculate the response function.

To implement this procedure we first determine $\bu$ and $\bv$ in terms
of the applied forces ${\bf f}^{\rm u,v}$.  We find 
\begin{equation}
\label{def_G}
{\cal G}_{ij}^{-1} \cdot \left( \begin{array}{c} u_j \\ v_j \end{array} 
\right) = \left( \begin{array}{c}  f^{{\rm u}}_j \\ \Pt_{jk} f^{{\rm v}}_k 
\end{array} \right). 
\end{equation}
The $6\times 6$ matrix ${\cal G}^{-1}$, described in detail in appendix 
\ref{append_G}, can be decomposed into a   
$2 \times 2$ matrix of $3 \times 3$ blocks.  In Eq.~(\ref{def_G}) the 
vectorial indices (shown) run over all three spatial directions while the 
indices labeling the $2 \times 2$ blocks (suppressed) run over the space 
spanned by the displacement field ($\bu$) and the fluid velocity field ($\bv$).
To determine $\bu$ and $\bv$ in terms of the applied forces on the 
medium, $\fu$, and $\fv$ we simply need to invert ${\cal G}^{-1}$.  
This inversion is easy owing
to the fact that the four $3 \times 3$ blocks of the matrix, displayed in 
Appendix \ref{append_G}, all mutually commute. 

We now integrate this result over all wavevectors, ${\bf k}$, to determine 
the fluctuating position (${\bf r}(\omega)$) and velocity 
(${\bf w}(\omega)$) of the point at the origin of the 
(unstressed) material. Because the problem is isotropic we find that the
vectorial part of each $2 \times 2$ block is particularly simple (see Appendix
\ref{append_G}): 
\begin{equation}
\label{def_g}
G^{\rm (n,m)}(\omega ) \delta_{ij} =  \int_{\left| {\bf k} \right| < 
k_{\rm max}} {\cal G}^{({\rm n, m})} _{ij}({\bf k}, \omega )
\frac{ d^3 k}{(2 \pi)^3} .
\end{equation}
The limits on the wavevector integral come from the cutoff imposed by the 
rigidity of the sphere, Eq.~(\ref{force-label}).  We have a pair of equations
specifying the position and velocity of the point at the origin of the two--fluid
medium (the position of the sphere) in terms of the force on the elastic 
network and the transverse part of the force on the fluid,
\begin{eqnarray} 
\label{rw_eqn1}
r_{i} (\omega ) &=& G^{{\rm (u,u)}}(\omega ) F^{{\rm u}}_i ( \omega ) + 
G^{{\rm(u,v)}}(\omega ) F^{{\rm v}}_i (\omega ) \\
\label{rw_eqn2}
w_i (\omega ) &=& G^{{\rm (v,u)}} (\omega ) F^{{\rm u}}_i ( \omega ) + 
G^{{\rm (v,v)}} (\omega ) F^{{\rm v}}_i (\omega ).
\end{eqnarray}
The boundary condition, Eq.~(\ref{boundary_con}),  becomes $ w_i(\omega) = -i 
\omega r_i(\omega)$. Imposing this condition on Eqs.(\ref{rw_eqn1}), 
(\ref{rw_eqn2}) fixes the ratio of 
${\bf F}^{{\rm u}}( \omega )$ to ${\bf F}^{{\rm v}}(\omega )$.  Using this
ratio  we write the displacement of the sphere, ${\bf r}(\omega )$, 
in terms of the 
total force that the sphere exerts on the medium, $-{\bf F}_{\rm b}(\omega )$:
\begin{equation}
\label{def_gamma}
{\bf r}(\omega ) = - \gamma(\omega ) {\bf F}_{\rm b}(\omega ), 
\end{equation}
where the function $\gamma(\omega )$ is given by
\begin{equation}
\label{gamma}
\gamma (\omega ) = \frac{ 1}{1 - X(\omega )} \left[ G^{{\rm (u,u)}}(\omega ) - 
G^{{\rm (u,v)}}(\omega ) X(\omega ) \right], 
\end{equation}
where the function $X(\omega )$ can be written in terms of the $G$'s as well:
\begin{equation}
\label{X_def}
X(\omega ) = \frac{ i \omega G^{{\rm (u,u)}}(\omega ) + G^{{\rm(v,u)}}
(\omega )}{ i \omega G^{{\rm (u,v)}}(\omega ) + G^{{\rm (v,v)}}(\omega ) }.
\end{equation}
Using Eq.~(\ref{def_gamma}) in Eq.~(\ref{Newt}) to eliminate ${\bf F}_{\rm b}$,
we find the position response of the bead to an applied force as defined in 
Eq.~(\ref{response_function}).
\begin{equation}
\label{solution}
\alpha^{-1}(\omega ) = \gamma^{-1}(\omega) - \omega^2 M.
\end{equation}
In essence Eqs.~(\ref{gamma}),(\ref{X_def}) and Eq.~(\ref{solution}) completely 
determine our solution for the response function. 
In order to study the detailed form of the response function, however, 
 we need to discuss the four $G^{\rm (n,m)}, ({\rm n,m} = {\rm u,v})$ 
as functions of frequency.  We begin this task in the next section by first
discussing each of the four $G$'s in turn and then looking in detail at the 
response function, $\alpha(\omega )$ which is made up of these $G$'s.

\section{The Response Function}
\label{response}

\begin{figure}
\epsfxsize=0.95\columnwidth
\centerline{\epsfbox{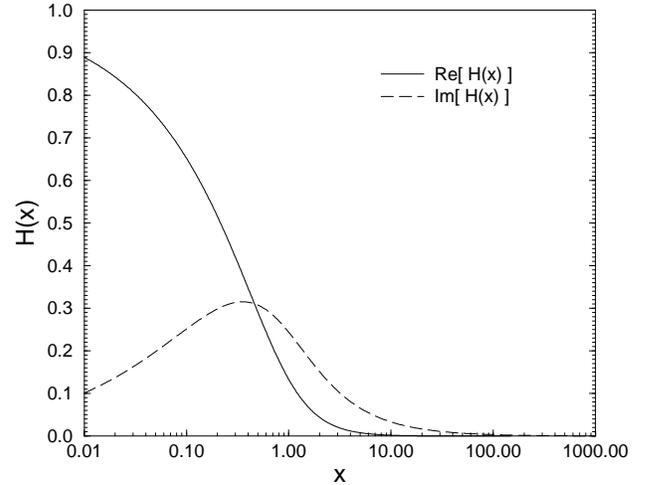}}
\vspace{0.1cm}
\caption[]{
The real and imaginary parts of the cross over function $H(x)$, which determines
the crossover from low--frequency ($\omega \ll \omega_{\rm B}$) behavior,
in which the network compressibility plays a role in the dynamics of the bead,
to the high--frequency ($\omega \gg \omega_{\rm B}$) incompressible dynamics.
In the response function $x$ is the ratio of the observation frequency to 
the frequency scale $\omega_{\rm B}$.  
}
\label{H_plot}
\end{figure}

We first look at the response of the elastic network to forces applied directly
to that network, $G^{{\rm(u,u)}}(\omega )$.  
We introduce the following notation:
the complex shear modulus of the two--fluid medium will be denoted by the 
usual $G(\omega) = \mu(\omega) - i \omega \eta$.  It should be noted that 
whereas $\eta$ represents the viscosity of the background solvent -- see 
our estimate of the drag force density coefficient $\Gamma$ -- the elastic
network may in general be viscoelastic. Its shear modulus will be 
given in general by a complex, frequency--dependent shear modulus, $\mu$.
We find that 
$G^{{\rm(u,u)}}(\omega )$ takes the form:
\begin{equation}
\label{G11}
G^{{\rm (u,u)}}(\omega ) = \frac{1}{ 6 \pi a G(\omega)} 
\left[ 1 + \frac{ G(\omega)}{
4 \mu + 2 \lambda} H \left( \frac{ \omega }{\omega_{\rm B}}\right)  + J(\omega)
\right] 
\end{equation}
where the function $H(x)$ is specified by the integral:
\begin{equation}
\label{H_def}
H(x) = 1 - \int_{0}^{1} \frac{ dz}{1 + \frac{ i z^2}{x}}.
\end{equation}
This function is plotted in figure \ref{H_plot}.
The function $J(\omega)$ is given by
\begin{equation}
\label{J_def}
J(\omega)= \int_{0}^{1} dz \frac{ \beta(\omega) + \Delta_0(z,\omega) z^2 
\left( 1 - \mu/G(\omega) \right) }{z^2 \left( 1 + \mu 
\Delta_0(z,\omega)/ G(\omega) \right)
- \beta(\omega)}. 
\end{equation}
The frequency scale, 
\begin{equation}
\label{omega_B}
\omega_{\rm B} = \frac{\left( 2 \mu + \lambda\right)}{\Gamma} k_{\rm max}^2,
\end{equation}
has also been introduced in Eq.~(\ref{G11}).
In addition to defining these two functions, we have introduced a set of new,
frequency--dependent parameters.  The first of these is
\begin{equation}
\label{beta_def}
\beta(\omega) = \frac{ \omega^2 \rF}{k_{\rm max}^2 G(\omega)} = \frac{ 4 a^2
\omega^2 \rF}{G(\omega) \pi^2},
\end{equation}
which measures the importance of the fluid inertia in determining the 
dynamics of the medium as may be noted by checking that this parameter 
can be expressed as the ratio of the sphere radius to the inertial 
decay length in the two--fluid medium\cite{Ferry:80}. 
We have also introduced
the parameter $ \Delta_0(z,\omega)$ defined by
\begin{equation}
\label{delta_def}
\Delta_0(z,\omega) = \frac{ -i \omega \rF + \eta k^2_{\rm max} z^2}{\Gamma}
 \approx -i \omega \xi^2/\nu + \left( z   k_{\rm max} \xi\right)^2,  
\end{equation}
which may be identified as the inverse viscous response function divided by
the drag coefficient $\Gamma$.  Using our estimate for the drag coefficient, we
may determine the magnitude of $\Delta_0$ under the conditions of a typical
microrheological experiment. 
The first term on the RHS of 
Eq.~(\ref{delta_def}) measures the ratio of the observation frequency to the
viscous dissipation rate at the length scale of the network mesh size $\xi  $.
We may estimate this latter time scale for typical experiments on 
actin\cite{Schnurr:97} to be in the $10$MHz range, well above any other 
frequencies of interest.  Thus $\omega \ll \nu/ \xi^2$.  Similarly the second 
term at the right of Eq.~(\ref{delta_def}) is small assuming 
that the sphere is much larger
than the mesh size, $a \gg \xi $.  We may reasonably set $\Delta_0$ to zero 
while discussing our result. 
The expression Eq.~(\ref{J_def}) may then be greatly
simplified by noting that $\Delta_0$ is vanishingly small under the typical 
conditions of a microrheological experiment.  
With this approximation we rewrite 
Eq.~(\ref{J_def}) as
\begin{equation}
\label{approx_J}
J(\omega) \approx \beta(\omega) \int_{0}^{1} dz \frac{ 1}{z^2 - \beta(\omega)}.
\end{equation}
It is now clear from this simplified, approximate form that $J(\omega)$ 
contains corrections to the response function coming from the
inertia of the two--fluid medium.  In direct analogy with the long--time 
tails in a Newtonian fluid (see Appendix \ref{Newtonian_fluid}), the 
lowest order in frequency inertial corrections coming from 
Eq.~(\ref{approx_J}) are of the
form $\sqrt{\beta(\omega)}$.  In a purely viscous medium this produces the 
standard $\omega^{1/2}$ corrections. In the viscoelastic medium, which we 
now study, the frequency dependence of the corrections will depend on the 
detailed form of the complex shear modulus.  We take up this point again in 
discussing the actin system in our conclusions.

Returning to the function $H$ introduced in Eq.~(\ref{G11}) and defined by
Eq.~(\ref{H_def}), we note that for large $x$, the function goes to zero as
$H(x) \sim i/(3 x)$.  See figure \ref{H_plot}.
For observation 
frequencies much larger than the frequency scale $\omega_{\rm B}$, the 
term proportional to $H(\omega/\omega_{\rm B})$ goes to zero, while for
frequencies much less than  $\omega_{\rm B}$, this term makes a finite 
correction to the response function.  The physical interpretation of the
$H$ function is made clear by recognizing that the crossover frequency 
is simply the  decay time of the network compression mode [whose 
dispersion relation is given in Eq.~(\ref{long-damped-incom})] at the 
length scale of the bead.  At frequencies much lower than $\omega_{\rm B}$,
the effect of the network compression mode upon the dynamics of the bead is
significant while at frequencies high compared to $\omega_{\rm B}$ the 
network is viscously locked to the incompressible fluid. Therefore, this
longitudinal mode of the network plays no role in the high--frequency bead 
dynamics.  The function $H(\omega/\omega_{\rm B})$ controls the 
cross-over from compressible network dynamics to incompressible network 
Finally we note that the zero--frequency response of the network to a localized
force on the network takes the Stokes mobility form, which is a standard 
result in the mechanics of elastic media.

Continuing our exploration of the response function, we consider the response
of the network to a force on the fluid, $G^{{\rm (u,v)}}$.  
We find that this term
has the form
\begin{equation}
\label{G12}
G^{{\rm (u,v)}}(\omega)= \frac{ 1}{6 \pi a G(\omega)} \int_{0}^{1} dz
\frac{z^2}{z^2 \left(  1 + \Delta_0(z,\omega) \frac{ \mu}{G(\omega)}\right)  
- \beta(\omega)}.
\end{equation}
Once again we may set $\Delta_0(z,\omega) = 0$ making only a small error in 
the combined limits: $\omega \ll \nu/\xi^2, k_{\rm max} \xi \ll 1$ so we 
may simplify the above expression to yield
\begin{equation}
\label{G12_simple}
G^{{\rm (u,v)}}(\omega) \simeq 1 + J(\omega),
\end{equation}
where $J(\omega)$ was defined in Eq.~(\ref{J_def}).  

We are now in a position
to see under what limiting conditions will the calculated response function
reduce to the GSER.  Examining $\alpha(\omega)$ at low enough frequencies so 
that we may ignore the bead inertia term ($- \omega^2 M$) we find the 
response function to be
\begin{eqnarray}
\alpha(\omega) &\simeq& \frac{1}{6 \pi a G(\omega)} \left[ 1 + 
J(\omega) + \right. \nonumber \\
\label{response2}
& & + \left. \frac{ 1}{1 - X(\omega)}  \frac{ G(\omega)}{4 \mu + 2 \lambda} 
H \left( \frac{ \omega}{\omega_{\rm B}} \right) \right].
\end{eqnarray}
For bead dynamics at frequency $\omega$ small enough so that we may
ignore the inertial corrections contained in $J(\omega)$, {\it i.e.\/}
$ \omega < \omega^\star, \beta(\omega^\star) = 1$, we may set $J =0$.  If, 
on the other hand, $\omega$ is much larger than $\omega_{\rm B}$ we may
ignore corrections to the bead's fluctuations coming from the thermal excitation
of the network compression mode and thus set $H(\omega /\omega_{\rm B}) = 0$.
Since we will be able to show that for typical values of the material parameters
(in actin solutions for example\cite{Schnurr:97}) $\omega_{\rm B} < 
\omega^\star$, there will exist a range of frequencies, $ \omega_{\rm  B} 
\longleftrightarrow \omega^\star$, for which the response function is 
well approximated by the GSER.  In order to discuss deviations from the GSER,
however, we need to study the form of $X(\omega)$.

We find that the function $X(\omega)$ is given by
\begin{eqnarray} 
X(\omega) &=& \frac{ i \omega \eta}{\mu \left( k_{\rm max} \xi\right)^2} \times \\
\nonumber
& & \left[ 
\int_{0}^{1} dz \frac{z^4}{z^2 \left( 1 + \frac{\mu}{G(\omega)} 
\Delta_0(z,\omega) \right) - \beta(\omega) } \right]^{-1} \times \\ \nonumber
&\times& \left\{ \int_{0}^{1} dz \frac{ z^2 \Delta_0(z,\omega)}{\left( 1 + 
\frac{ \mu}{G(\omega)} \Delta_0(z,\omega) \right) z^2 - \beta(\omega)} \right.
\\ \nonumber
 &+& 
\left. \frac{ G(\omega)}{2 \lambda + 4 \mu} 
H \left( \frac{ \omega}{\omega_{\rm B}} \right) \right\}
\label{X}
\end{eqnarray}
It is important to note that the prefactor multiplying $X(\omega)$ above 
contains $(k_{\rm max} \xi )^{-2}$ which we have argued is typically quite 
large.  The large number is, however, multiplied by the ratio of the 
viscous stress in the background fluid to the stress in the viscoelastic 
network.  For experimentally realizable frequencies this ratio is quite small.
Thus $X(\omega)$ presents a small correction to the response function in 
a majority of interesting cases.  We develop this point further in the 
discussion of these result presented in the next section.

\section{Summary}
\label{conclusions}

The response function of the rigid, spherical probe particle in the two--fluid
medium has been calculated, see Eqs.~(\ref{solution}),
(\ref{response2}).  Through a detailed study of the position response function
of the probe particle embedded in a two--fluid, viscoelastic medium to an 
externally applied force, we have checked the validity of the GSER.  Our
results show that there exits a frequency range, $ \omega_{\rm B} 
\longleftrightarrow \omega^\star, \left| \beta(\omega^\star) \right| = 1$,
over which the GSER is a good approximation to the full response function.  
We now consider recent experiments on actin solutions to determine the 
width of this regime, and if it is possible experimentally detect the 
deviations from GSER behavior.  

In recent actin experiments\cite{Schnurr:97}, the shear modulus was found to be 
well approximated over a frequency range extending from about $10$Hz (above the
plateau frequency) to the highest measured frequencies of a few KHz by:
\begin{equation}
\label{experimental}
G'_{\rm exp}(\nu) \simeq  G''_{\rm exp}(\nu) \simeq 
\left( \frac{ \nu  }{  ({\rm Hz})  }  \right)^{3/4} 10 
\frac{ {\rm dynes}  }{   {\rm cm}^2   }.
\end{equation}
The $\nu^{3/4}$ frequency dependence is in agreement with the theories of 
a number of groups\cite{Amblard:96,threefourstheory}. 
The shear modulus of the network dominates over the shear viscosity in the 
background fluid up to very high frequencies.  This may be checked by comparing
$\left|G'_{\rm exp}(\omega)\right| \simeq  \left|G''_{\rm exp}(\omega)\right|
$ to the viscous shear stress modulus: $\omega \eta$, taking the viscosity
to be that of water.  Using Eq.~(\ref{experimental}) we can compare the 
relative magnitudes of the shear stress in the the viscoelastic network 
to that of the background fluid ($\omega \eta $).  Clearly at large enough
frequencies the fluid will carry the larger part of the stress in the material,
while below some crossover frequency, the network shear modulus is the 
the dominant contributor to the mechanical properties of 
the two--fluid material. A simple calculation shows that this crossover 
frequency is approximately $6 \times 10^8 {\rm Hz}$, which is well above all 
experimentally accessible frequencies so that the network shear modulus 
is always the principle contributor to the two--fluid shear modulus.
It is this dominance of the network contribution to the shear modulus in 
the material that 
allows us to ignore corrections coming from $X(\omega)$ in our solution of 
the response function in Eq.~(\ref{response2}).
We now check whether inertial effects are important in these measurements 
which ranged up to frequencies of a few kHz.  There are two sources of 
inertial effects: those coming from the fluid inertial and those coming from
the mass of the probe particle.  We first look at the fluid inertia.  Using
our expression for $\beta(\nu)$ in Eq.~(\ref{beta_def}) we find the 
cross--over frequency, $\nu^\star, \beta(\nu^\star) = 1$ to be given by
\begin{equation}
\label{nu_star}
\nu^\star = \left( 1.6  a^2 \right)^ {-4/5} {\rm Hz}
\end{equation}
where $a$ is measured in centimeters.  The experiment employed probe sphere
sizes ranging from one to five microns yielding cross--over frequencies in 
the range of $1.7$MHz --- $131$kHz, which, given that these experiments 
probe frequencies up to only a few kHz, suggests that the onset of inertial 
effects should be unobservable at present.  
It should be remembered, however,  that the cross--over to the inertial 
regime is slow, being governed by $\sqrt{\beta(\omega)}$ so the effects of 
fluid inertia may be detectable at significantly lower frequencies.  
Nevertheless, we do 
not believe that the present experiments are probing the fluid inertial regime.
There is a similar inertial effect due to the mass of the probe particle. To
determine the frequency onset of the signature of the probe particle's 
inertia, we may compare the particle inertial term $- \omega^2 M$ to the 
dominant contribution to the response function at high frequency, the 
generalized Stokes mobility of sphere, $ 6 \pi G(\omega) a$.  This comparison
gives roughly the same estimates as from the fluid inertia estimate above. The
similarity of the two estimates is not surprising since the probe particle 
is of nearly the same density as the fluid. 

We note from  appendix \ref{rheology} that the effect of inertia is not
negliable in the traditional rheological measurements of soft 
materials\cite{Tanner}.  We find that if the soft material is probed using a
standard parallel--plate shear cell, inertial corrections to the response
function, $G(\omega) = \mu - i \omega \eta$, enter at frequencies 
such that the oscillating plates excite shear waves in the medium whose 
decay length is shorter than the plate separation $L$. In the limit that 
the plate separation is much larger than the mesh size of the network 
(a necessary assumption for the application of our continuum theory) 
these inertial corrections may be expressed in terms of a plate separation 
independent scaling factor. In this limit, the experimentally 
determined response function
$G_{\rm exp} (\omega)$ is related to the expected, low--frequency response 
function $G(\omega ) = \mu - i \omega \eta$ by the relation
\begin{equation}
\label{rheology_ans}
G_{\rm exp}(\omega) = G(\omega ) y \coth(y)
\end{equation}
where the dimensionless variable $y$ is defined in terms of the plate 
separation $L$, and the shear wave speed $c$ and damping rate $\Delta$ --- 
see Eq.~(\ref{dispersion-shear}) and Eqs.~(\ref{c}),(\ref{Delta}) in 
appendix \ref{rheology}:
\begin{equation}
y = \left( i \frac{ \omega }{c} - \frac{ \omega^2}{c^3} \Delta\right) L. 
\end{equation}
In the low--frequency limit such that $\left|y \right|  \ll 1 $ it 
is clear that this expression reduces to the expected result: 
$G_{\rm exp}(\omega) \simeq G(\omega)$.  For the type of parallel--plate 
experiment under discussion, a sample thickness of one millimeter implies that 
the inertial corrections can be neglected for frequencies below $10$kHz.  
This upper bound on frequencies imposed by the appearance of inertial 
corrections is of the same order of magnitude as the analogous bound 
determined for the microrheology response function\cite{Levine:00}.

We now turn to the determination of the low--frequency limit of the 
GSER relation.  The lower bound of this frequency range is given by 
$\omega_B$.  Using 
our expression for $\omega_{\rm B}$ given in Eq.~(\ref{omega_B}) we find that
the low--frequency cross--over to the network  compression 
regime occurs at
\begin{equation}
\nu_{\rm B} \simeq \frac{ 2 \mu + \lambda}{\eta} \frac{ \pi}{8} \left( \frac{ %
\xi }{a} \right)^2.
\end{equation}
Given typical material parameters for entangled actin solutions, taking the 
elastic moduli to be on the order of the plateau modulus and taking the 
network mesh size to be on the order of a tenth of a micron we find that 
$\nu_{\rm B} \simeq 1$Hz.  This is on the order of the plateau frequency and
is certainly probed by experiment.  

To summarize our work we note that the response function probed by a single
particle, microrheological experiment contains information about {\it all\/}
of the hydrodynamic modes of the system.  In other words the fluctuations of
the probe particle are in response to all the thermally excited modes of 
the system, whereas in a standard, macrorheological experiment, one explicitly
determines the response of the system to an externally applied shear strain.
If the medium admits hydrodynamic modes that are not
simply shear waves, the microrheological response function 
can not be expressed entirely in terms of the material's complex
shear modulus as determined from standard rheology.  On the other hand, if the
hydrodynamic modes of the medium are simply shear waves then we expect that 
the simple correspondence between microrheological and standard rheological 
measurements, as expressed by the GSER will hold at low enough frequencies.  At
higher frequencies, both techniques will encounter the inertial effects.
Microrheology, however,
allows the exploration of the mechanical response of the medium at much higher
frequencies than those probed by standard rheology, so the importance of the
inertia of both the medium and the probe particle itself cannot be overlooked
{\it a priori\/}.

For the model viscoelastic medium which we have studied there is an extra
hydrodynamic mode (as compared to an incompressible, viscous fluid) which 
introduces of lower frequency bound on the validity of the GSER.  This lower
bound has some experimental significance for entangled actin solutions as this
lower bound occurs near to the frequency of the rubber plateau in this material.
The inertial effects, however, should not be relevant to current experiments
that study the high frequency, single chain dynamics of the system.

We would like to thank J.C.~Crocker for communicating unpublished results and 
many useful discussions.   We would also like to thank R.D.~Kamien, 
F.C.~MacKintosh, D.C.~Morse, and A.G.~Yodh for helpful discussions. This work 
was supported in part by the NSF MRSEC Program under grant No. DMR96--32598.

\appendix

\section{The response of a sphere in a Newtonian fluid}
\label{Newtonian_fluid}

In this appendix we test our approximate solution method by calculating the 
response function of a sphere in a Newtonian fluid. This problem has a 
well--known solution\cite{Landau} which lets us check the validity of our 
approximation scheme.  As throughout this article, we assume that the spherical
particle undergoes simple harmonic motion of an amplitude small compared to 
its size so that we may neglect the flow advection term in the Navier--Stokes
equation even in the high Reynolds number limit.  The problem we wish to 
solve is simply stated: What is the force acting the bead if it is observed
to undergo simple harmonic motion of the form: $\bv = {\rm Re}\left[ v_0 %
\exp \left( -i \omega t \right) \right] $? 

The motion of the spherical particle (of mass $M$)
obeys Newton's second law:
\begin{equation}
\label{append_Newton}
M \dot{\bv} = {\bf F} + {\bf F}_{\rm b}
\end{equation}
where ${\bf F}$ is the externally applied force on the bead and 
${\bf F}_{\rm b}$ is the force due to the fluid acting on the bead. We will 
use our approximation scheme to calculate that force, ${\bf F}_{\rm b}$.  First
we solve for the velocity field of the fluid given that some force 
${\bf F}_{\rm v}(\bx,t) = {\bf F}_{\rm v}(\bx, \omega ) \exp(-i \omega t)$ 
is applied to it using  
\begin{equation}
\rF \dot{\bv} = \eta \nabla^2 \bv - {\bf \nabla } P + {\bf F}_{\rm v}(\bx,t).
\end{equation}
Additionally, we require the incompressibility of the fluid:
\begin{equation}
{\bf \nabla } \cdot \bv = 0.
\end{equation}
We calculate the fluid velocity at the origin (the location of the bead) by
integrating over all wavevectors, $\bk$,
\begin{equation}
v_\alpha( \bx = 0, \omega ) = \int \frac{ d^3 k}{(2 \pi )^3} 
\frac{ \Pt_{ \alpha \beta }(\bk) {\bf F}_{\rm v}(\bk,\omega)}
{-i \omega \rF + \eta k^2}
\end{equation}

Noting the spherical symmetry of the bead, we demand that 
${\bf F}_{\rm v}(\bk,\omega )= {\bf F}(\omega) {\cal F}(k)$
is a function of the magnitude of $k$ alone, allowing us to perform the angular
integrations above and leading to the following result for $\gamma_{\rm v}$
defined by
\begin{equation}
\bv( \bx = 0, \omega ) = \gamma_{\rm v}(\omega ) {\bf F}(\omega ).
\end{equation}
We determine
\begin{eqnarray}
\label{gammav}
\gamma_{\rm v}(\omega ) &=& \frac{ 1}{6 \pi \eta a } \left[ \frac{ 2}{\pi }
\int_{0}^{\infty} d z {\cal F}(z) \right.\\ \nonumber 
 &+& \left. \frac{ 2 i}{\pi }
\frac{ \omega }{\omega_\nu} \int_{0}^{\infty} d z {\cal F}(z) \frac{ 1}{z^2 -
i \omega/\omega_{\nu }} \right]
\end{eqnarray}
where $z = k a$, the frequency scale $\omega_{\nu } = \nu a^{-2}$ 
(where $\nu = \eta/\rF$ is the kinematic viscosity) is the viscous dissipation
rate at the length scale of the sphere, and the as yet 
unknown function ${\cal F}$ is determined by  the 
$k$ dependence of ${\bf F}_{\rm v}(\left|\bk\right|,\omega )$, or, in other
words, how we localize the force of the bead upon the fluid at the surface 
of the bead.  One clear choice is to localize the force on the interface of 
the fluid and the sphere: $ {\bf F}_{\rm v}(\bx) = \frac{ F}{2 \pi a} \delta %
\left( \left|x\right|^2 - a^2 \right) $ where $F$ total force exerted by 
the sphere on the fluid and we have suppressed the oscillatory time dependence.
An even simpler choice, which we have made throughout the paper, is localize 
the force in wavevector space via: ${\bf F}_{\rm v}(\bk) = \Theta \left( 
\frac{ \pi }{2 a} - \left| k \right| \right) $.  The first choice leads to 
${\cal F}(z) = \frac{ \sin(z)}{z}$
while the second version yields ${\cal F}(z) = \frac{ \pi}{2} \Theta( 1 - z)$.
Hereafter we refer to the first version as the ``shell localization'' and the 
second version as the ``volume localization''.

Using the shell localization we find that Eq.~(\ref{gammav}) simplifies to
the exact expression:
\begin{equation}
\label{answer_append1}
\gamma_{\rm v}^{-1}(\omega ) = 6 \pi \eta a  \exp \left[ \left( 1-i 
\right) \sqrt{ \frac{ \omega }{2 \omega_{\nu }}}\, \right] .
\end{equation}
We will be concerned only with the expansion of the above expression for 
$ \omega \ll  \omega_{\nu }$.  Using the volume localization, on the 
other hand, we find that the exact result to all orders is more complicated
but to order $\sqrt{\omega / \omega_{\nu }}$ we find an identical result to 
that above (Eq.~(\ref{answer_append1})),
\begin{equation}
\gamma_{\rm v}^{-1}(\omega ) = 6 \pi \eta a \left[ 1 +  \left( 1 - i 
\right) \sqrt{ \frac{ \omega }{2 \omega_{\nu }}} + {\cal O}\left(%
 \frac{ \omega }{ \omega_{\nu }} \right) \right]. 
\end{equation}
Of course, to this order in frequency, we may ignore the inertial of the bead
and from Eq.~(\ref{append_Newton}) we note that $\gamma_{\rm v}^{-1}(\omega )$
is then identical to the inverse response function we sought. This result
agrees with the standard solution of this problem arrived at through the 
complete solution of the boundary value problem\cite{Landau} to the order
in frequency shown above.  At higher orders in frequency, starting with 
${\cal O}(\omega / \omega_{ \nu})$ where the bead's inertia comes into play, 
deviations between our approximate 
calculation of the fluid's inertia and the exact result appear.  Our result
over estimates the ${\cal O}(\omega / \omega_{ \nu})$ contribution to the 
fluid inertial by a factor of about $5.5$.
Based on this analysis we expect
similar accuracy in the two--fluid calculations using the volume localization
scheme that are presented in this paper.  As is discussed in the conclusions,
the inaccuracy of our results at high frequencies is not relevant to the 
current set microrheological measurements. These experiments have not 
yet probed the transition to the inertial regime, which should, in fact, be
 well described by our (correct) order ${\cal O}(\sqrt{\omega / \omega_{ \nu}})$
fluid inertia terms.

\section{The \protect{${\cal G}^{-1}$} Matrix}
\label{append_G}
Here we write the matrix ${\cal G}^{-1}$ in its  $2 \times 2$ block form.  
We introduce the viscous response function in the fluid: 
$ \Delta^{-1}(\bk, \omega ) = -i \omega \rF + \eta k^2$, and elastic response
of the network, with additional damping due to the coupling to the viscous 
fluid, decomposed into its transverse, $D^{-1}_{\rm T}(\bk, \omega ) =
- \omega^2 \rho + \mu k^2 -i \omega \Gamma $, and longitudinal, 
$D^{-1}_{\rm L}(\bk, \omega ) =
- \omega^2 \rho + \left( 2 \mu + \lambda \right)  k^2 -i \omega \Gamma $, 
parts.  In terms of these functions we may write ${\cal G}^{-1}$ as
\begin{equation}
\label{inverseG}
\left( \begin{array}{ccc} D^{-1}_{\rm T}
(\bk, \omega )\Pt_{ij} + D^{-1}_{\rm L}
(\bk, \omega )\Pl_{ij} & -\Gamma \delta_{ij} \\
i \omega \Gamma \Pt_{ij} & \Delta^{-1}(\bk, \omega ) \delta_{ij} + 
\Gamma \Pt_{ij}
\end{array}
\right).
\end{equation}
All four $3 \times 3$ blocks shown above are proportional to either the 
identity matrix or the transverse or longitudinal projectors.  Since all
three of these matrices are mutually commuting we see that the inversion
of ${\cal G}^{-1}$ is quite simple. 

After performing this matrix inversion we find that the four $2 \times 2$ 
blocks are given by:
\begin{eqnarray}
\label{block_one}
{\cal G}^{(1,1)}_{ij} &=& \frac{ \left( \Delta^{-1}(\bk,\omega) + 
\Gamma \right) \Pt_{ij}(\bk)
}{ D^{-1}_{\rm T}(\bk, \omega ) \left( \Delta^{-1}(\bk,\omega) + \Gamma \right)
+ i \omega \Gamma^2 } + \\ 
\nonumber  &+& \frac{ \Pl_{ij}(\bk)}{-\omega^2 \rho + ( 2 \mu + 
\lambda) k^2} \\ 
{\cal G}^{(1,2)}_{ij} &=& \frac{ \Gamma \Pt_{ij}(\bk)}{\left( \Delta^{-1}(\bk,
\omega) + \Gamma \right)  D^{-1}_{\rm T}(\bk, \omega ) + i \omega \Gamma^2} \\
{\cal G}^{(2,1)}_{ij} &=& \frac{ -i \omega \Gamma \Pt_{ij}(\bk)}{\left( 
\Delta^{-1}(\bk, \omega) + \Gamma \right)  D^{-1}_{\rm T}(\bk, \omega ) + 
i \omega \Gamma^2} \\
{\cal G}^{(2,2)}_{ij} &=& \frac{ D^{-1}_{\rm T}(\bk, \omega ) 
\Pt_{ij}(\bk)}{\left( \Delta^{-1}(\bk, \omega) + \Gamma \right)  
D^{-1}_{\rm T}(\bk, \omega ) + i \omega \Gamma^2}
\end{eqnarray}
In the first of the above equations we have found the response of the network
to a force on the network.  The second (and third) of the response functions
shown above gives the response of the network (fluid) to a force on the fluid
(network).  The final response function is the response of the fluid to 
a force on the fluid.  This interpretation becomes clear in the decoupled 
limit where $\Gamma \longrightarrow 0$. Here the fluid and the network do not
interact so ${\cal G}^{(1,2)} = {\cal G}^{(2,1)} = 0$.  The response of the 
network to forces on the network is given by
\begin{equation}
\lim_{\Gamma \longrightarrow 0} {\cal G}^{(1,1)}_{ij} = \frac{\Pt_{ij}(\bk)}{
-\omega^2 \rho + \mu k^2} + \frac{ \Pl_{ij}(\bk)}{-\omega^2 \rho + (2 \mu
+ \lambda) k^2},
\end{equation}
showing the standard transverse (first term) and longitudinal (second term) 
response of an isotropic, elastic medium to an applied force.  The response
of the fluid to a force on a fluid is similarly in accord with basic 
hydrodynamics,
\begin{equation}
\lim_{\Gamma \longrightarrow 0} {\cal G}^{(2,2)}_{ij} = \frac{\Pt{ij}(\bk)}{
-i \omega \rF + \eta k^2}.
\end{equation}

\section{Standard rheology on the two--fliud medium}
\label{rheology}
In this appendix we calculate the response function of the medium to an
externally applied shear strain in order to predict the result of a traditional 
rheological measurement.  We check this result in order to compare it with the
response of the probe particle discussed in this article.  We do not expect
to see any evidence of the longitudinal network mass density mode since 
the system will be subjected to a pure shear strain by moving the boundaries
of the material.  Nevertheless we do expect to observe departures of the 
shear response from the simple value of $G(\omega) = \mu - i \omega \eta $ 
due to inertial terms.  We will thus compare the effect of inertial in the 
standard rheological experiment to the microrheological experiment via this 
calculation.

We begin with the equations of motion defining the two--fluid medium, 
Eqs.~(\ref{ueom}) -- (\ref{incompress}).  We now consider a slab of this 
composite material held between two, parallel, rigid plates normal to the 
$\hat{z}$ axis located at $z = 0,L$.  The
slab is unbounded in the $xy$ plane.  In order to calculate the complex 
shear response of the material we will move the top plate ($z=L$) harmonically,
$ \bu (z=L) = \hat{x} U_0 e^{-i \omega t }$ 
while holding the bottom plate fixed.
Given these boundary conditions we calculate the required shear stress on 
the top plate, $\sigma(z= L)_{xz}$.  The ratio of this shear stress to the 
imposed shear strain $U_0/L$ is the complex shear complex shear modulus at
the frequency $\omega$.  

By the translational invariance of the problem in the $xy$ plane we may restrict
our search for the resulting network displacement and fluid velocity fields 
to those of the form:
\begin{eqnarray}
\label{uguess}
\bu &=& F(z) e^{-i \omega t} \hat{x} \\
\label{vguess}
\bv &=& G(z) e^{-i \omega t} \hat{x}. 
\end{eqnarray}
Using the incompressibility of the fluid we find that $P$, the hydrostatic
pressure is an harmonic function.  Since the pressures at both plates are
equal, $P$ is constant.  Using Eqs.~(\ref{ueom}),(\ref{veom}) 
we find two coupled, ordinary differential equations for $F$ and $G$:
\begin{eqnarray}
\label{diff1}
\mu \partial_z^2 F + \rho \omega^2 F + i \omega \Gamma F + \Gamma G &=& 0 \\
\label{diff2}
\eta \partial_z^2 G + i \omega \rF G - \Gamma G - i \omega \Gamma F &=& 0
\end{eqnarray}
Putting in $ F = F_i e^{\lambda_i z}, G = G_i e^{\lambda_i z}$, we find that
nontrivial solutions of the above differential equations can only exist for
values of the $\lambda$ that solve the characteristic equation,
\begin{eqnarray}
\nonumber
\eta \mu \lambda^4 + \lambda^2 \left[ - \mu \Gamma + i \omega \left( \rF \mu
+ \eta \Gamma \right) + \eta \rho \omega^2 \right] + \\
\label{characteristic}
 + \left[ - \omega^2
\left( \rho + \rF \right) \Gamma + i \omega^3 \rF \rho \right] &=& 0.
\end{eqnarray}
The above equation has four roots coming in two pairs of roots having the 
same absolue value, {\it i.e.\/} $\lambda_1, \ldots, \lambda_4$ where 
$\lambda_1^2 = \lambda_2^2$ and $\lambda_3^2 = \lambda_4^2$.  Corresponding
to these four eigenvalues there are four eigenvectors of the form: 
$ \left( F_i, \gamma_i F_i \right)$.  It should be noted that, since the 
eigenvector equation depends only upon the square of the eigenvalue, the 
coefficients $\gamma_i$, $i = 1, \ldots, 4$ have the following relations:
$ \gamma_1 = \gamma_2$ and $\gamma_3 = \gamma_4$.  We now may write the 
general solution to  Eqs.~(\ref{diff1}), (\ref{diff2}) as a linear superposition
of the four eigenvectors discussed above:
\begin{eqnarray}
\label{ans1}
F(z) &=& \sum_{i=1}^4 F_i e^{\lambda_i z} \\
G(z) &=& \sum_{i=1}^4 \gamma_i F_i e^{\lambda_i z}.
\end{eqnarray}
We have four boundary conditions to determine the remaining four constants. At
the bottom plate we require stick boundary conditions for the fluid and 
the network at that immobile plate: 
\begin{equation}
G(0) = F(0) = 0.
\end{equation}
We also impose stick boundary conditions at the harmonically oscillating, 
upper plate:
\begin{eqnarray}
G(L) &=& -i \omega U_0 \\
F(l) &=& U_0.
\end{eqnarray}

Using the above boundary conditions we determine the network displacement
field to be
\begin{eqnarray}
\nonumber
F(z) &=& \frac{ U_0}{\lambda_1^2 - \lambda_3^2} \left[ \lambda_1^2 \frac{ 
\sinh(\lambda_3 z)}{\sinh(\lambda_3 L)} - \lambda_3^2 \frac{
\sinh(\lambda_1 z)}{\sinh(\lambda_1 L)} \right] + \\
& & + \frac{ U_0}{\lambda_1^2 - \lambda_3^2} \frac{ \rho \omega^2}{\mu} 
\left[ \frac{\sinh(\lambda_3 z)}{\sinh(\lambda_3 L)} - 
\frac{\sinh(\lambda_1 z)}{\sinh(\lambda_1 L)} \right], 
\end{eqnarray}
and the fluid velocity field to be
\begin{eqnarray}
\nonumber
G(z) &=& \frac{ -i \omega U_0}{\lambda_1^2 - \lambda_3^2} \left[ 
\lambda_1^2 \frac{ \sinh(\lambda_3 z)}{\sinh(\lambda_3 L)} - \lambda_3^2 
\frac{ \sinh(\lambda_1 z)}{\sinh(\lambda_1 L)} \right] + \\
& & +  \frac{ -i \omega U_0}{\lambda_1^2 - \lambda_3^2}  \frac{ i \omega \rF}{
\eta} \left[ \frac{\sinh(\lambda_3 z)}{\sinh(\lambda_3 L)} -
\frac{\sinh(\lambda_1 z)}{\sinh(\lambda_1 L)} \right].
\end{eqnarray}

We may now calculate the complex shear modulus that would be measured by
a standard rheology experiment performed on our two--fluid medium.  We calculate
the applied stress stress divided by the applied shear strain to get the 
response function, $G_{\rm exp}(\omega)$. 
\begin{equation}
G_{\rm exp}(\omega) = \left( \mu \left. \frac{ d F}{d z} \right|_{z=L} + \eta 
\left. \frac{ d G}{d z} \right|_{z =L} \right) \frac{ L}{U_0}.
\end{equation}
After some minor rearrangements we arrive at
\begin{equation}
G_{\rm exp}(\omega) = \left[ \left( \mu - i \omega \eta \right) + 
\frac{ \omega^2 \left( \rho + \rF \right)}{\lambda_1 \lambda_3} \right] J(L).
\end{equation}
In the above equation, the term in the parentheses is the expected result
for the response function.  It is simply the sum of the complex shear response
of the network and the viscous response of the permeating fluid.  The second
term in the brackets is clearly an inertial correction to this standard result.
Both of these terms are multiplied by a system size dependent scaling factor,
$J(L)$.  This function when expressed in terms of the dimensionless 
variables, $x = L \lambda_1$, $y = L \lambda_3$, takes the form 
\begin{equation}
\label{Jdef}
J(L) = \frac{ x  y}{x^2 - y^2} \left[ x \coth(y) - y \coth(x) \right].
\end{equation}

We expect that approximation: $ G_{\rm exp} \simeq \mu - i \omega \eta $ should
hold at least as the limiting behavior of $ G_{\rm exp}(\omega) $ at 
low frequencies.  To check this we need to consider the frequency dependence
of the two eigenvalues appearing above: $\lambda_1$ and $\lambda_3$.

In the limit of low frequency we find that these roots of the characteristic
polynomial, Eq.~(\ref{characteristic}) take the form:
\begin{eqnarray}
\label{lam1}
\lambda_1^2 &=& \frac{ \Gamma}{\eta} - i \omega \left( \frac{ \rF \mu + \eta 
\Gamma }{\mu \eta} \right)  + \frac{ \omega^2 \rF}{\mu } + {\cal O}(\omega^3)\\
\label{lam3}
\lambda_3^2 &=& - \frac{ \omega^2 \left( \rho + \rF \right) }{\mu} - 
i \omega^3 \left( \frac{ \rF + \rho}{\mu}\right)^2 \times \\
\nonumber
&\times & \left[ \frac{ \mu 
\rF^2}{\left( \rho + \rF \right)^2 \Gamma } + \frac{ \eta }{\rho + \rF} \right]
+ {\cal O}(\omega^4).
\end{eqnarray}
We note that $\lambda_1$ at low frequency is the inverse of a microscopic 
length since $\Gamma/\eta \sim \xi^{-2}$.  In a macroscopic shear experiment of
the type we are currently considering, the plate separation is much larger 
than this microscopic length: $L \lambda_1 \gg 1$.  In Eq.~(\ref{Jdef}) we
may take $ x =  L \lambda_1 \gg 1$.  If we now take the modulus of $y$ to be
small, $\left|y \right|= L \left| \lambda_3 \right| \ll 1$
we find that $J(L)$ does, in fact, reduce to unity.  The second limit 
is valid for low frequencies.  To satisfy this inequality, both the imaginary
and real parts of $y$ must be small.  We consider the physical implications 
of these two conditions independently.  

It may be checked by comparing Eq.~(\ref{lam3})
to Eq.~(\ref{dispersion-shear}) that to lowest order in frequency,  $\Re(y) = 
\frac{ \Delta \omega^2}{c^2} \left( \frac{ L}{c} \right)$ where $c$ is the 
transverse shear wave speed and $\Delta$ is the tranverse shear wave damping 
rate as given in Eq.~(\ref{dispersion-shear}), 
\begin{eqnarray}
\label{c}
c &=& \sqrt{\frac{ \mu}{\rho + \rF}} \\
\label{Delta}
\Delta &=& \frac{ 1}{2} \left[  \frac{ \mu
\rF^2}{\left( \rho + \rF \right)^2 \Gamma } + \frac{ \eta }{\rho + \rF} \right].
\end{eqnarray}
The imaginary part of $y$, on the other hand, takes the form: 
$\Im(y) = - L \omega / c$. 
Requiring that the modulus of $y$ to be small (and thus requiring
both the imaginary and real parts of $y$ to be small) is equivalent to 
demanding that the sample be subjected to a uniform (affine) shear 
deformation as we discuss below.  The standard intrepretation of
a macroscopic--shear rheological experiment supposes that the sample has 
been affinely deformed by the imposed shear.  The validity of that assumption
is essential if one is to determine
the complex shear modulus from a macroscopic--shear rheological experiment 
where the applied stress and resultant strain are measured only at the 
sample boundaries.

First, the condition that $\Re(y) \ll 1$ implies that the damping rate of the 
shear waves multiplied by the shear wave travel time across the sample must be 
small.  In other words the shear wave in the two--fluid composite medium should
not be appreciably damped on the length scale of the sample thickness. 
If, on the other hand, the sample thickness is much larger than the
shear penetration depth, $\delta$,  only the small portion of the sample 
of thickness equal to that penetration depth is strained.  The shear strain 
in the sample that results from the applied stress is not  uniformly 
$U_0/L$ but rather a spatially dependent quantity. It is of order $U_0/\delta$ 
within one penetration depth of the moving plate and essentially zero throughout
the remaining depth of the sample.  

Second, the condition that the imaginary part of $y$ be small requires that 
the oscillation frequency of the plate be smaller than the inverse shear wave 
propagation time across the thickness of the sample.  If the applied shearing
frequency is too high, there is a significant time lag between the imposed 
displacement at the sample boundary and the resulting deformation of the 
material at points far from that boundary.  The result of that lag is 
once again
to produce a nonuniform shear displacement in the sample so that the shear 
strain is not simply $U_0/L$ but rather some more complicated function of 
position in the medium.  This conclusion can be simply checked for a purely 
elastic, one--component medium.  For the same reasons as discussed above, the 
measurement of the applied shear stress at the boundary will not 
result in an accurate determination of the shear modulus of the material. 

Finally, we give the complex shear response to first order in frequency as 
measured in a traditional rheology experiment,
\begin{eqnarray} 
\label{G_approx}
G_{\rm exp}(\omega) &=& G(\omega) + i \omega c \left( \rho + \rF \right) 
\sqrt{\frac{\eta}{\Gamma}} + \\
\nonumber
 & & + \omega^2 \left[ \frac{ \Delta}{c} 
\left( \rho + \rF \right) \sqrt{\frac{\eta}{\Gamma}} - G(\omega) 
\frac{ L^2}{c^2} \right] 
\end{eqnarray}
Here we have written the expected complex shear response as $G(\omega) = \mu 
- i \omega \eta$ and have made use of $c$ and $\Delta$ defined in 
Eqs.~(\ref{c}), (\ref{Delta}). In the above expression, 
Eq.~(\ref{G_approx}), we have not made use of the inequality 
$\Gamma L^2/\eta \sim \left( L/\xi  \right)^2 \gg 1$ to further simplify 
the result.


\begin{references}
\bibitem{active-passive}
Hereafter we refer explicitly only to the passive mode of measurement.  
with no loss of generality since fundamental equilibrium statistical physics 
links these results to the active measurements.
\bibitem{lowphi}
Reconstituted F-actin networks solutions with network volume fractions
less than $0.1\%$ can be strongly entangled.  See reference 
\cite{Schnurr:97}.
\end{references}
\end{document}